\documentclass[11pt]{llncs}

\usepackage{xspace}
\usepackage[english]{babel}
\usepackage[latin1]{inputenc}
\usepackage{float}
\usepackage{subfigure}

\usepackage[svgnames, table]{xcolor}

\usepackage{a4wide}

\usepackage{amsmath,amssymb}

\usepackage{graphicx}
\usepackage[export]{adjustbox}
\usepackage{comment}
\usepackage{textcomp}
\usepackage{booktabs}

\usepackage[hyphens]{url}
\usepackage[breaklinks=true,pdftex]{hyperref}  

\def\ie{{i.e.},~}
\def\eg{{e.g.},~}

\def\z3{{\sc Z3}\xspace}

\def\hash{\mathbf{hash}}
\def\zero{\mathbf{zero}}
\newcommand{\figref}[1]{\figurename~\ref{#1}}

\colorlet{vert}{green!70!black}
\colorlet{rouge}{red!70!black}
\colorlet{orange}{orange!100!black}
\colorlet{bleu}{cyan!80!white!80!black}
\colorlet{gris}{black!10!white}

\usepackage{tikz}
\usetikzlibrary{arrows,backgrounds,calc,trees}

\pgfdeclarelayer{background}
\pgfsetlayers{background,main}

\newcommand{\convexpath}[3]{
[   
    create hullnodes/.code={
        \global\edef\namelist{#1}
        \foreach [count=\counter] \nodename in \namelist {
            \global\edef\numberofnodes{\counter}
            \node [xshift=#3] at (\nodename) [draw=none,name=hullnode\counter] {};
        }
        \node at (hullnode\numberofnodes) [name=hullnode0,draw=none] {};
        \pgfmathtruncatemacro\lastnumber{\numberofnodes+1}
        \node at (hullnode1) [name=hullnode\lastnumber,draw=none] {};
    },
    create hullnodes
]
($(hullnode1)!#2!-90:(hullnode0)$)

\foreach [
    evaluate=\currentnode as \previousnode using \currentnode-1,
    evaluate=\currentnode as \nextnode using \currentnode+1
    ] \currentnode in {1,...,\numberofnodes} {
  let
    \p1 = ($(hullnode\currentnode)!#2!-90:(hullnode\previousnode)$),
    \p2 = ($(hullnode\currentnode)!#2!90:(hullnode\nextnode)$),
    \p3 = ($(\p1) - (hullnode\currentnode)$),
    \n1 = {atan2(\y3,\x3)},
    \p4 = ($(\p2) - (hullnode\currentnode)$),
    \n2 = {atan2(\y4,\x4)},
    \n{delta} = {-Mod(\n1-\n2,360)}
  in 
    {-- (\p1) arc[start angle=\n1, delta angle=\n{delta}, radius=#2] -- (\p2)}
}
-- cycle
}

\tikzset{
  inode/.style = {align=center, inner sep=0pt, text centered,
    font=\sffamily, circle, draw=white, black, fill=white, 
    text width=2.1em, very  thick}
  }

\newcounter{mynote}
\newlength\mynotewidth
\usepackage{listings}

\lstdefinelanguage{dafny}{
  sensitive=true,
  keywords={},
  otherkeywords={
  <,>, <=, >=, |, ==, :=, int, seq, tail, init, last, first, take
  },
  basicstyle=\fontsize{9}{11}\selectfont\sffamily,
  keywords = [2]{var, function, lemma, method, ghost, if, then, else, ensures, requires, decreases, while, do, return, od, assert, invariant},
  keywordstyle={\bfseries\color{orange}},
  keywordstyle=[2]{\bfseries\color{blue!80!black}},
  identifierstyle=\color{black},
  comment=[l]{//},
  commentstyle=\color{gray}\ttfamily,
  stringstyle=\color{red}\ttfamily,
  morestring=[b]',
  morestring=[b]",
  frame=lines
}

\AtBeginDocument{}

\title{\LARGE \bf Verification of the Incremental Merkle Tree Algorithm with Dafny}

\author{Franck Cassez
  \institute{
    ConsenSys\\
    \email{franck.cassez@consensys.net}
  }
}

\makeatletter
\renewcommand{\paragraph}{%
  \@startsection{paragraph}{4}%
  {\z@}{0.9ex \@plus 1ex \@minus .2ex}{-1em}%
  {\normalfont\normalsize\bfseries}%
}
\makeatother

\begin{document}

\pagestyle{plain}
\maketitle

\thispagestyle{empty}

\begin{abstract}
The Deposit Smart Contract (DSC) is an instrumental component of the 
Ethereum 2.0 Phase 0 infrastructure.
We have developed the first machine-checkable version of the incremental Merkle tree algorithm used in the  DSC.
We present our new and original correctness proof of the algorithm along with the Dafny machine-checkable version.
The main results are: 1) a new proof of total correctness; 2) a software artefact with the proof in the form of the complete Dafny code base and  3) new provably correct optimisations  of the algorithm.
\end{abstract}

\section{Introduction}\label{sec-intro}

Blockchain-based decentralised platforms process transactions between parties and record them in an immutable distributed ledger.
Those platforms were once limited to handle simple transactions   
but the next generation of  platforms will  routinely run \emph{decentralised applications} (DApps) that enable users to make complex transactions (sell a car, a house or more broadly, swap assets) without the need for an institutional or governmental trusted third-party. 

\paragraph{\bf \itshape  Smart Contracts.}

More precisely, the transactions are \emph{programmatically} performed by \emph{programs} called \emph{smart contracts}. 
If there are real advantages having smart contracts act as third-parties to process transactions, there are also lots of risks that are inherent to computer programs: they can contain \emph{bugs}. 
Bugs can trigger runtime errors like \emph{division by zero} or \emph{array-out-of-bounds}. 
In a networked environment these types of vulnerabilities can be exploited by malicious attackers over the network to disrupt or take control of the computer system.
Other types of bugs can also compromise the business logic of a system, \eg an implementation may contain subtle errors (\eg using a \verb|+=| operator in C instead of \verb|=+|) that make them deviate from the initial intended specifications.

Unfortunately it is extremely hard to guarantee that programs and henceforth smart contracts implement the correct business logics, that they are free of common runtime errors, or that they never run into a non-terminating computation.\footnote{In the Ethereum ecosystem, programs can only use a limited amount of resources, determined by the \emph{gas limit}. So one could argue that non-terminating computations are not problematic as they cannot arise: when the gas limit is reached a computation is aborted and has no side effects. It follows that a non-terminating computation (say an infinite loop due to a programming error) combined with a finite gas limit will abort and will result in the system being unable to successfully process some or all \emph{valid} transactions  and this is  a serious issue.}
There are notorious examples of smart contract vulnerabilities that have been exploited and publicly reported:
in  2016, a \emph{reentrance} vulnerability in the Decentralised Autonomous Organisation (DAO) smart contract was exploited to steal more than USD50 Million. There may be several non officially reported similar attacks that have resulted in the loss of assets.

\paragraph{\bf \itshape The Deposit Smart Contract in Ethereum 2.0.}

The next generation of Ethe\-reum-based networks, Ethereum 2.0, features 
a new \emph{proof-of-stake} consen\-sus protocol.
Instead of miners used in Ethereum 1.x, the new protocol relies on \emph{va\-lidators} to create and
\emph{validate}  blocks of transactions that are added to the led\-ger. 
The protocol is designed to be fault-tolerant to up to $1/3$ of Byzantine (\ie malicious or dishonest) validators.  
To discourage validators to deviate from an honest behaviour, they have to \emph{stake} some assets in  Ether (a crypto-currency), and if they are dishonest they can be 
\emph{slashed} and lose (part of) their stake. The process of staking is handled by the \emph{Deposit Smart Contract (DSC)}:
a validator sends a transaction (``stake some Ether'') by \emph{calling} the DSC.
The DSC has a \emph{state} and can update/record the history of deposits that have occurred so far.

As a result the DSC is a mission-critical
component of Ethereum 2.0, and any errors/crashes could result in inaccurate tracking of the deposits or downtime which in turn may compromise the integrity/availability of the whole system. 

This could be mitigated if the DSC  was a  simple piece of code, 
but, for performance reasons, it relies on sophisticated data structures and algorithms to
maintain the list of deposits so that they can be communicated over the network efficiently:
the history of deposits is summarised as a unique number, a \emph{hash}, computed using a \emph{Merkle} (or \emph{Hash}) tree. 
The tree is built incrementally using the \emph{incremental Merkle tree algorithm}, and
as stated in~\cite{deposit-cav-2020}:
\begin{quote}
    \em ``The efficient incremental algorithm leads to the DSC implementation being unintuitive, and makes it non-trivial to ensure its correctness.'' 
\end{quote}

\paragraph{\bf \itshape Related Work.}

In this context, it is not surprising that substantial efforts, au\-di\-ting, review~\cite{suhabe-dsc}, testing and formal verification~\cite{formal-inc-merkle-rv,deposit-cav-2020} has been invested to guarantee the reliability and integrity (\eg resilience to potential attacks) of the DSC.
The DSC has been the focus of an end-to-end analysis~\cite{deposit-cav-2020}, including the bytecode\footnote{A limitation is that the bytecode is proved using a  non-trusted manual specification.} that is executed on the Ethereum Virtual Machine (EVM). 
However, the \emph{incremental Merkle tree algorithm} has not been \emph{mechanically verified} yet, even though a pen and paper  proof has been proposed~\cite{formal-inc-merkle-rv} and \emph{partially} mechanised using the K-framework~\cite{k-2019}.
An example of the limitations of the mechanised part of the proof in~\cite{formal-inc-merkle-rv} is that it does not contain a formal (K-)de\-finition of Merkle trees.  
The mechanised sections (lemmas 7 and 9) pertain to some invariants of the algorithm 
but not to a proper correctness specification based on Merkle trees.   
The K-framework and KEVM, the formalisation of the EVM in K, has been used to analyse a number of other smart contracts~\cite{rv-sc}.   
There are several techniques and tools\footnote{\url{https://github.com/leonardoalt/ethereum_formal_verification_overview}.} \eg ~\cite{solc-vstte-19,fmbc-20,DBLP:conf/cpp/AmaniBBS18,csi-mythx,harvey}, for auditing and analysing smart contracts writ\-ten in Solidity (a popular language to write Ethereum smart contracts) or EVM bytecode, but they offer limited capabilities to verify complex functional requirements.

Interesting properties of incremental Merkle trees were established in~\cite{ogawa} using the MONA prover. 
This work does not prove the algorithms in the DSC which are 
designed to minimise gas consumption and hence split into parts: insert a value in a tree, and compute the root hash.
Moreover, some key lemmas in the proofs could not be discharged by MONA. 


%

\smallskip 

The gold standard in program correctness is a complete logical proof
that can be \emph{mechanically checked} by a prover.
This is the problem we address in this paper: to design a \emph{machine-checkable proof} for the DSC algorithms (not the bytecode) using the Dafny language and verifier.
The DSC has been deployed in Nov\-ember 2020.
To the best of our knowledge, our analysis, completed in October 2020, 
provided the first fully mechanised proof 
that the code logic was correct, and free of runtime errors.
There seem to be few comparable case-studies of Dafny-verified (or  other verification-aware programming languages like Whiley~\cite{whiley-setss-2018}) code bases. The most notorious and complex one is probably the IronFleet/IronClad~\cite{ironfleet-2015} dis\-tri\-bu\-ted system, along with some non-trivial algorithms like DPPL~\cite{dpll-dafny-19} or Red-Black trees~\cite{DBLP:journals/jar/Pena20}, or operating systems, FreeRTOS scheduler~\cite{matias_program_2014}, and ExpressOS~\cite{DBLP:conf/asplos/MaiPXKM13}.  
Other proof assistants like Coq~\cite{Paulin-Mohring2012}, Isabelle/HOL~\cite{Nipkow-Paulson-Wenzel:2002} or Lean~\cite{DBLP:conf/cade/MouraKADR15} have also been extensively used to write machine-checkable proofs of algorithms~\cite{lammich-ijcar-2020,lammich-mc-ta,timsort-jdk,ghc-mergesort} and software systems~\cite{DBLP:conf/sosp/KleinEHACDEEKNSTW09,DBLP:journals/jar/Leroy09}. 


\paragraph{\bf \itshape Our Contribution.}
We present a thorough analysis of the incremental Merkle tree algorithm used in the DSC. 
Our results are available as software artefacts, written using the CAV-awarded Dafny\footnote{\url{https://github.com/dafny-lang/dafny}} verifi\-ca\-tion-aware programming language~\cite{dafny-ieee-2017}.
This provides a self-contained machine checkable and reproducible proof of the DSC algorithms. 
Our contribution is many-fold and includes:
\begin{itemize}
    \item a \emph{new original simple proof} of the incremental Merkle tree algorithm.
    In contrast to the previous non-mechanised proof in~\cite{formal-inc-merkle-rv}  we do not attempt to directly prove the existing algorithm, but rather to \emph{design} and refine it.
    Our proof is \emph{parametric} in the height of the tree, and 
         \emph{hash} functions; 
    \item a \emph{logical specification} using a formal definition of Merkle trees, and a \emph{new functional version} of the algorithm that is proved correct against this specification; the functional version is  used to specify the invariants for the proof of the imperative original version~\cite{vitalik-merkle} of the algorithm; 
    \item a repository\footnote{\url{https://github.com/ConsenSys/deposit-sc-dafny}} with the complete Dafny source code of the specification, the algorithms and the proofs, and comprehensive documentation;
    \item some new provably correct simplifications/optimisations; 
    \item some reflections on the practicality of using a verification-aware programming language like Dafny
    and some lessons learned from this experience.
\end{itemize}

\vspace{-1.0em}

\section{Incremental Merkle Trees}\label{sec-example}

\paragraph{\bf \itshape Merkle Trees.}
A \emph{complete  (or perfect) binary tree} is such that each non-leaf node has exactly two children, and the two children have the same \emph{height}.
An example of a complete binary tree is given in \figref{fig-compute-root-hash}. 
A \emph{Merkle (or hash) tree} is a complete binary tree the nodes of which are decorated
with \emph{hashes} (fixed-size bit-vectors). The hash values of the leaves are given and the hash values of the internal (non-leaf) nodes are computed by \emph{combining} the values of their children with a binary function $\hash$.
It follows that a Merkle tree is a complete binary tree decorated with a \emph{synthesised attribute} defined by a binary function. 

Merkle trees are often used in distributed ledger systems to define a \emph{property} of a collection of elements \eg a list $L$ of values. 
This property can then be used \emph{instead of the collection itself} to verify,\footnote{More precisely the verification result holds with high probability as the chosen hashing functions may (rarely) generate collisions.}
 using  a  mechanism called \emph{Merkle proofs}, that 
data received from a node in the distributed system is not corrupted.
This is a crucial optimisation as the size of the collection is usually large (typically up to $2^{32}$) and using a compact representation is instrumental to obtain time and space efficient communication and a reasonable transactions' processing throughput.
\begin{quote}
    \it In this work, we are not concerned with Merkle proofs but rather with the (efficient) computation of the $\hash$ attribute on a Merkle tree.
\end{quote}
The actual function used to compute the values of the internal nodes is not re\-le\-vant in the incremental Merkle tree algorithms' functional logics and without loss of generality we may treat it as a parameter \ie a given  binary function.\footnote{In the code base, the $\hash$ function is uninterpreted and its type is generic.}
In the sequel we assume that the decorations of the nodes are integers, and we use in the examples a simple function $\hash : \text{Int} \times \text{Int} \longrightarrow \text{Int}$ defined by $\hash(x, y) = x - y - 1$ instead of an actual (\eg \texttt{sha256}-based) hash function.

\paragraph{\bf \itshape Properties of Lists with Merkle Trees.}

A  complete binary tree of height\footnote{The height is the length of the longest path from the root to any leaf.} $h$ has $2^h$ leaves and $2^{h + 1} - 1$ nodes.
Given a list $L$ of integers (type \text{Int}) of size $|L| = 2^h$, we let $T(L)$ be the Merkle tree for $L$: the values of the leaves of $T(L)$, from left to right, are the elements of $L$ and $T(L)$ is attributed with the $\hash$ function. 
The value of the attribute at the root of $T(L)$, the \emph{root hash},
defines a property of the list $L$.
It is straightforward to extend this de\-fi\-ni\-tion to lists $L$ of size  $|L| \leq 2^h$ by right-padding the list with \emph{zeroes} (or any other default values.) 
Given a list $L$ of size $|L| \leq 2^h$, let $\overline{L}$ denote $L$ right-padded with $2^h - |L|$ default values. The Merkle tree associated with $L$ is $T(\overline{L})$, and the  root hash of $L$ is the root hash of $T(\overline{L})$.
Computing the  root hash of a tree $T(\overline{L})$ requires to traverse all the nodes of the tree and thus is \emph{exponential} in the height of the tree.

\paragraph{\bf \itshape The Incremental Merkle Tree Problem.}
A typical use case of a Merkle tree in the context of Ethereum 2.0 is to represent properties of lists that \emph{grow monotonically}. 
In the DSC, 
a Merkle tree is used to record the list of validators and their stakes or deposits.
A compact representation of this list, as the root hash of a Merkle tree, is com\-mu\-nicated to the nodes in the network rather than the tree (or list) itself.
However, as  mentioned before, each time a new deposit is appended to the list, computing the new root hash using a standard synthesised-attribute computation algorithm requires exponential time in $h$.
This is clearly impractical in a distributed system like Ethereum in which the height of the tree is $32$ and the number of nodes is $2^{33} - 1$.

Given (a tree height) $h > 0$, $L$ a list with $|L| < 2^h$, and $e$ a new element to add to $L$, 
the incremental Merkle tree problem (IMTP) is defined as follows:\footnote{Polynomial in the height of the tree $h$. The operator $+$ is list concatenation.}
\begin{quote}
    \it 
    Can we find $\alpha(L)$ a \textbf{polynomial-space abstraction} of $T(L)$ such that we can compute in 
    \textbf{polynomial-time}: 1) the root hash of $T(L)$ from $\alpha(L)$, and 2) the abstraction $\alpha(L + [e])$ from $\alpha(L)$ and $e$? 
\end{quote}
Linear-time/space algorithms 
to solve the IMTP were originally pro\-po\-sed by V. Buterin in~\cite{vitalik-merkle}.
However, the correctness of these algorithms is not obvious.
In the next section, we analyse the IMTP,
and we present the main properties that en\-able us to \emph{design}
polynomial-time recursive algorithms and to \emph{verify} them.

\vspace{-1em}
\section{Recursive Incremental Merkle Tree Algorithm}\label{sec-3}


In this section we present the main ideas of the \emph{recursive} algorithms to insert a new value in a Merkle tree and to compute the new root hash (after a new value in inserted) by re-using (\emph{dynamic programming}) previously computed results.
%

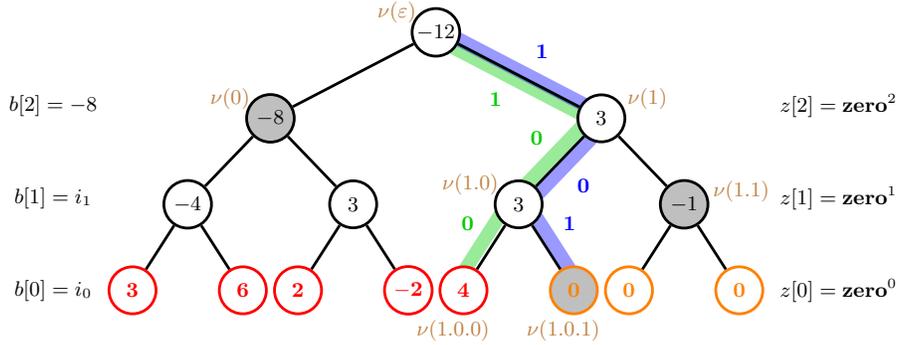
\begin{figure}[tbhtp]
    \centering
    \scalebox{0.88}{
      \begin{tikzpicture}[-,>=stealth',level/.style={sibling distance = 5cm/#1,
        level distance = 1.3cm, very thick}] 
        \node [inode,label={[brown]left,yshift=0.3cm,xshift=.2cm:$\nu(\varepsilon)$}] (n14) {$-12$}
            child { 
                node [inode,fill=lightgray,label={[brown]left,yshift=0.3cm,xshift=.2cm:$\nu(0)$}] (n12) {$-8$}  
                child {   
                node [inode] (n8) {$-4$} 
                child { 
                    node [inode,draw=red] (n0) {\textcolor{red}{$\mathbf 3$}}
                }
                child { 
                  node [inode,draw=red] (n1) {\textcolor{red}{$\mathbf 6$}}
                }
                }
                child { 
                node [inode] (n9) {$3$}
                child { 
                    node [inode,draw=red] (n2) {\textcolor{red}{$\mathbf 2$}}
                }
                child { 
                    node [inode,draw=red] (n3) {\textcolor{red}{$\mathbf{-2}$}}
                    }
                }
            }  
            child { 
                node [inode,label={[brown]left,yshift=0.3cm,xshift=1.5cm:$\nu(1)$}] (n13) {$3$}  
                child { 
                    node [inode,label={[brown]left,yshift=0.3cm,xshift=.2cm:$\nu(1.0)$}] (n10) {$3$}
                    child { 
                    node [inode,draw=red,label={[brown]left,yshift=-0.6cm,xshift=0.9cm:$\nu(1.0.0)$}] (n4) {\textcolor{red}{$\mathbf 4$}}
                    edge from parent node[label={[green!80!black]above left:$\mathbf 0$}] {}
                    } 
                    child { 
                    node [inode,draw=orange,fill=lightgray,label={[brown]left,yshift=-0.6cm,xshift=0.9cm:$\nu(1.0.1)$}] (n5) {\textcolor{orange}{$\mathbf 0$}}
                    edge from parent node[label={[blue]above right:$\mathbf 1$}] {}
                    }
                    edge from parent node[label={[blue]below right:$\mathbf 0$},label={[green!80!black]above left:$\mathbf 0$}] {}
                } 
                child { 
                    node [inode,fill=lightgray,label={[brown]left,yshift=0.2cm,xshift=1.8cm:$\nu(1.1)$}] (n11) {$-1$}
                    child { 
                    node [inode,draw=orange] (n6) {\textcolor{orange}{$\mathbf 0$}}
                    }
                    child { 
                    node [inode,draw=orange] (n7) {\textcolor{orange}{$\mathbf 0$}}
                    }
                } 
                edge from parent node[label={[blue]above right:$\mathbf 1$},label={[green!80!black]below left:$\mathbf 1$}] {}
            };
        \begin{pgfonlayer}{background}
        \begin{scope}            
            \fill[green!80!black,opacity=0.4] \convexpath{n14, n13, n10, n4, n10, n13, n14}{3pt}{-5pt};
        \end{scope}
        \fill[blue,opacity=0.4] \convexpath{n14, n13, n10, n5, n10, n13, n14}{3pt}{4pt};
        \end{pgfonlayer}
    
        \node [left of=n0,xshift=-.2cm] (b2) {$b[0] = i_0$};
        \node [above of=b2, yshift=.4cm] (b1) {$b[1]= i_1$};
        \node [above of=b1, yshift=.4cm] (b0) {$b[2] = -8$};

        \node [right of=n7, xshift=.5cm] (z2) {$z[0] = \zero^0$};
        \node [above of=z2, yshift=.4cm] (z1) {$z[1] = \zero^1$};
        \node [above of=z1, yshift=.4cm] (z0) {$z[2] = \zero^2$};

        \end{tikzpicture}
    }
    \caption{A Merkle tree of height $3$ for list $L_2 = [3, 6, 2, -2, 4]$ and $\hash(x, y) = x - y -1$. The green path $\pi_1$ is encoded as $1.0.0$ (from root to leaf) and the blue path $\pi_2$ as $1.0.1$. 
    The left and right siblings of $\pi_1$ are shaded. The values of the right siblings of $\pi_1$ at levels $0$ and $1$ are $z[0]=\zero^0 = 0$ and $z[1]=\zero^1=\hash(0,0) = -1$. $i_0$ and $i_1$ are arbitrary values.}
    \label{fig-compute-root-hash}
    \end{figure}

\paragraph{\bf \itshape Notations.}
A \emph{path} $\pi$ from the root of a tree to a node can be defined as a se\-quence of bits (left or right) in $\{0, 1\}^*$.
In a Merkle tree of height $h$, the \emph{length}, $|\pi|$, of $\pi$ is at most $h$.
$\nu(\pi)$ is the \emph{node} at the end of $\pi$.
If $|\pi| = h$ then $\nu(\pi)$ is a leaf. 
 For instance $\nu(\varepsilon)$ is the root
of the tree, $\nu(0)$ in \figref{fig-compute-root-hash} is the node carrying the value $-8$ and $\nu(1.0.0)$ is a leaf.
The \emph{right sibling of a left node} of the form $\nu(\pi.0)$ is the node $\nu(\pi.1)$.
Left siblings are defined symmetrically.
A node in a Merkle tree is associated with a \emph{level} which is the distance from the node to a leaf in the tree. Leaves are at level $0$ and the root is at level $h$.
In a Merkle tree, level $0$ has $2^h$ leaves that can be indexed  left to right from $0$ to $2^h -1$.
The \emph{$n$-th leaf} of the tree for $0 \leq n < 2^h$ is the leaf at index $n$.

\noindent\begin{minipage}{\linewidth}
  \begin{lstlisting}[language=dafny,caption=Recursive Algorithm to Compute the Root Hash., captionpos=t, label={algo-compute-root}]
  computeRootUp(p:seq<bit>,left:seq<int>,right:seq<int>,seed:int):int
    requires |p| == |left| == |right| //  vectors have the same sizes
    decreases p 
  {
    if |p| == 0 then seed 
    else if last(p) == 0 then // node at end of p is a left node
      computeRootUp(init(p),init(left),init(right),hash(seed,last(right)))
    else // node at end of p is a right node
      computeRootUp(init(p),init(left),init(right),hash(last(left),seed))
  }
\end{lstlisting}
\end{minipage}
  
\paragraph{\bf \itshape Computation of the Root Hash on a Path.}
We first show that the root hash can be computed if we know the values of the \emph{siblings} of
the nodes on \emph{any} path, and the value at the end of the path.
For instance, 
If we know the values of the left and right siblings (shaded nodes) of the nodes
on $\pi_1$ (green path in 
\figref{fig-compute-root-hash}), and the value at the end of $\pi_1$, we can compute the root hash of the tree 
by propagating upwards the attribute $\hash$.
The value of the $\hash$ attribute at $\nu(1.0)$ is $\hash(4,\nu(1.0.1)) = 3$, at $\nu(1)$ it is 
$\hash(3, \nu(1.1)) = 3$ and at the root $\hash(\nu(0), \nu(1)) = \hash(-8, 3) = -12$. 

Algorithm  \texttt{computeRootUp} (Listing~\ref{algo-compute-root}) computes\footnote{For $l = l' + x$, $\mathtt{last}(l) = x$, $\mathtt{init}(l) = l'$, and for $l = x + l'$, $\mathtt{first}(l) = x$, $\mathtt{tail}(l) = l'$.} bottom-up in time linear in $|\mathtt{p}|$ the root hash
with \texttt{left} the list of values of the left siblings (top-down) on a path \texttt{p} (top-down), \texttt{right} the values of the right siblings (top-down) and \texttt{seed} the value at $\nu(\mathtt{p})$.
The generic version (uninterpreted hash) of the algorithm is provided in the \href{https://github.com/ConsenSys/deposit-sc-dafny/blob/50b16f96021368a839f932b5d666729405a305b0/src/dafny/smart/synthattribute/ComputeRootPath.dfy#L57}{\tt ComputeRootPath.dfy} file.

%
For the green path $\mathtt{pi}_1 = [1,0,0]$ in~\figref{fig-compute-root-hash}, 
$\mathtt{left} = [-8, i_1, i_0]$,  $\mathtt{right} = [-1, -1, 0]$ and the seed is $4$.
The evaluation of \texttt{computeRootUp} returns $-12$.

\smallskip 

Given a path $\pi$, if the leaves on the right of $\nu(\pi)$ all have the default value $0$, 
the values of the right siblings on the path $\pi$
only depend on the \emph{level} of the sibling in the tree. 
For example, the leaves on the right of $\pi_1$ (orange in \figref{fig-compute-root-hash}) all have the default value $0$.
The root hash of a tree in which all the leaves have the same default value only depends on the level of the root: $0$ at level $0$, $\hash(0, 0)$ at level $1$,
$\hash(\hash(0, 0), \hash(0, 0))$ at level  $2$ and so on.
Let $\zero^l$ be defined by: $\zero^l = 0 \text{ if $l = 0$ else $\hash(\zero_0^{l -1}, \zero_0^{l -1})$}$.

\vspace*{-0.5em}
\begin{quote}\em
  Given a path $\pi$, if all the leaves on the right of $\nu(\pi)$ have the default value, 
  any right sibling value at level $l$ on $\pi$ is equal to $\zero^l$.
\end{quote}
As an example in \figref{fig-compute-root-hash}, the right siblings on $\pi_1 = 1.0.0$ have values $0$ at level $0$, node $\nu(1.0.1)$,  and $\hash(0,0) = \zero^1 = -1$ at level $1$, node $\nu(1.1)$. 
\vspace*{-0.2em}
%
%
If a path \texttt{p} leads to a node with the default value $0$ and all the leaves right of $\nu(\mathtt{p})$ have the default value $0$, the root hash depends only 
on the values of the \texttt{left} and default \texttt{right} siblings.
Hence the root hash can be obtained by \texttt{computeRootUp(p, left, right, 0)}.
For the path $\mathtt{pi2} = [1, 0, 1]$ (\figref{fig-compute-root-hash}),
$\mathtt{left} = [-8, i_1, 4]$, $\mathtt{right} = [-1, -1, 0]$, 
\texttt{computeRootUp(pi2, left, right, 0)} returns $-12$.

As a result,   to compute the root hash of a tree $T(\overline{L})$, we can use a compact abstraction $\alpha(L)$ of $T(\overline{L})$ composed of  
the left siblings vector $b$ and the right siblings default values $z$ (\figref{fig-compute-root-hash}) of the path to the $|L|$-th leaf in $T(\overline{L})$.

\paragraph{\bf \itshape Insertion: Update the Left Siblings.}
Assume $\pi_1$ is a path to the $n$-th leaf and $ n < 2^h - 1$ (not the last leaf), 
 where the next value $v$ is to be inserted.
As we have shown before, if we have $b_1$ holding the values of left siblings of $\pi_1$, $z$ and $v$,
we can compute the new attribute values of the nodes on $\pi_1$ and the new root hash after $v$ is inserted.
Let $\pi_2$ be the path to the $n+1$-th leaf. 
If we can compute the  
values $b_2$ of the left siblings of $\pi_2$ as a function of $b_1$, $z$ and $v$, we have an efficient algorithm to
\emph{incrementally} compute the root hash of a Merkle tree: we keep track of the values of the left siblings $b$ on the path to the next available leaf, and iterate this process each time a new value is inserted.

As  $\nu(\pi_1)$ is not the last leaf, $\pi_1$ must contain at least one $0$, and
 has the form\footnote{$x^k, x \in \{0,1\}$ denotes the sequence of $k$ $x$'s.}  $\pi_1 = w.0.1^k$ with $w \in \{0, 1\}^*, k \geq 0$.
Hence, the path $\pi_2$ to the $n+1$-th leaf 
is $w.1.0^k$, 
arithmetically $\pi_2 = \pi_1 + 1$. An example of two consecutive paths is given in 
\figref{fig-compute-root-hash} with $\pi_1$ (green) and $\pi_2$ (blue) to the leaves at indices $4$ and $5$.

The related forms of $\pi_1$ (a path) and $\pi_2$ (the successor path) are useful to figure out how to incrementally compute the left siblings vector $b_2$ for $\pi_2$:
\begin{itemize}
  \item as the initial prefix $w$ is the same in $\pi_1$ and $\pi_2$, the values of the left siblings on the nodes of $w$ are the same in $b_1$ and $b_2$;
  \item all the nodes in the suffix $0^k$ of $\pi_2$ are left nodes and have right siblings.
  It follows that the corresponding $k$ values in $b_2$ are irrelevant as they correspond to right siblings, and we can re-use the corresponding $b_1$ values;
  \item hence $b_2$ is equal to $b_1$ except possibly for the level of the node at $\nu(w.0)$.
\end{itemize}
We now illustrate how to compute the new value in the vector $b_2$ on the example of 
\figref{fig-compute-root-hash}.
Let $\pi_1 = w.0$ and $\pi_2 = w.1$ with $w = 1.0$ and $|w| = 2$. 
For the top levels $2$ and $1$, $b_2$ is the same as $b_1$: $b_2[2] = b_1[2] = -8$ and
$b_2[1] = b_1[1] = i_1$.
For level $0$, the level of the node $\nu(w.0)$, the value at $\nu(w.0) = \nu(1.0.0)$ becomes the left sibling of the node $\nu(1.0.1)$ on $\pi_2$ at this level. So the new value of the left sibling on $\pi_2$ is exactly the new value, $4$, of the node $\nu(1.0.0)$ after $4$ is inserted.  

More generally, when computing the new root hash bottom-up on $\pi_1$, the first time we encounter a left node, at level $d$, we update the corresponding value of $b$  with the computed value of the attribute on $\pi_1$ at level $d$.
Algorithm\footnote{$+$ stands for list concatenation.} \texttt{insertValue} in Listing~\ref{algo-compute-sib} computes, in linear-time, the list of values of the left siblings (top-down) of the path $\mathtt{p} + 1$ using as input the list (top-down) of values left (resp. right) siblings 
\texttt{left} (resp. \texttt{right}) of \texttt{p} and \texttt{seed} the new value inserted at $\nu(\mathtt{p})$.
The generic (non-interpreted hash) algorithm is provided in the \href{https://github.com/ConsenSys/deposit-sc-dafny/blob/50b16f96021368a839f932b5d666729405a305b0/src/dafny/smart/paths/NextPathInCompleteTreesLemmas.dfy#L101}{\tt NextPathInCompleteTreesLemmas.dfy} file.

\noindent\begin{minipage}{\linewidth}
  \begin{lstlisting}[language=dafny,caption=Recursive Algorithm to Compute the New Left Siblings., captionpos=t, label={algo-compute-sib}]
insertValue(p:seq<bit>,left:seq<int>,right:seq<int>,seed:int):seq<int>
  requires |left| == |right| == |p| >= 1
  decreases p 
{
  if |p| == 1 then  // note that first(p) == last(p) in this case
    if first(p) == 0 then [seed] else left 
  else if last(p) == 0 then // we encounter a left node. Stop recursion.
      init(left) + [seed]
  else //  right node,move up on the path.
      insertValue(init(p),init(left),init(right),hash(last(left),seed)) 
        + [last(left)]
} 
\end{lstlisting}
\end{minipage}

We illustrate how the algorithm \texttt{insertValue} works with the example of \figref{fig-compute-root-hash}.
Assume we insert the seed $4$ at the end of the (green) path \texttt{pi1 = [1,0,0]}.
The left (resp. right) siblings' values  are given by $\mathtt{left} = [-8, i_1, i_0]$ (resp. $\mathtt{right} = [-1, -1, 0]$).
\texttt{insertValue} computes the values of the left siblings on the (blue) path \texttt{pi2 = [1,0,1]} after $4$ is inserted at the end of $\pi_1$: the first call terminates the algorithm 
and returns $[-8, i_1, 4]$ which is the list of left siblings that are needed on $\pi_2$.

In the next section we describe how to verify the recursive algorithms and the versions implemented in the DSC.

\vspace{-1em}
\section{Verification of the Algorithms}

In order to verify the implemented (imperative style/Solidity) versions of the algorithms of the DSC, we first 
prove total correctness of the recursive versions (Section~\ref{sec-3}) and them use them 
to prove the code implemented in the DSC.

\begin{quote}\em
  In this section, the Dafny code has been simplified and sometimes even altered 
  while retaining the main features, for the sake of clarity. 
  The code in this section may not compile. We provide the links to the files with the 
  full code in the text and refer the reader to those files.   
\end{quote}


\paragraph{\bf \itshape Correctness Specification.}


The (partial) correctness of our algorithms reduces to
checking that they compute the same values as the ones obtained 
with a synthesised attribute on a Merkle tree.
We have specified the data types \texttt{Tree}, \texttt{MerkleTree} and \texttt{CompleteTrees} and the relation between Merkle trees and lists of values (see \href{https://github.com/ConsenSys/deposit-sc-dafny/blob/50b16f96021368a839f932b5d666729405a305b0/src/dafny/smart/paths/NextPathInCompleteTreesLemmas.dfy#L101https://github.com/ConsenSys/deposit-sc-dafny/tree/50b16f96021368a839f932b5d666729405a305b0/src/dafny/smart/trees}{\tt trees} folder.)

The root hash of a \texttt{MerkleTree} \texttt{t} is \texttt{t.rootv}. 
The  (specification) function \texttt{buildMerkle(h, L, $\hash$)} returns a \texttt{MerkleTree} of height \texttt{h}, the  leaves of which are given by the values (right-padded) $\overline{\mathtt{L}}$, and the values on the internal nodes agree with the definition of the synthesised attribute $\hash$, \ie 
what we previously defined in Section~\ref{sec-example}  as $T(\overline{\texttt{L}})$.
It follows that \texttt{buildMerkle(h, L, $\hash$).rootv} is the root hash of  a Merkle tree with leaves $\overline{\mathtt{L}}$.

\paragraph{\bf \itshape Total Correctness.}
The total correctness proof for the \texttt{computeRootUp} function amounts to showing that 1) the algorithm always terminates and 2) the result of the computation is the same as the hash of the root of the tree. 
In Dafny, to prove termination, we need to provide a ranking function (strictly decreasing and bounded from below.) 
The length of the path \texttt{p} is a suitable ranking function 
(see the \texttt{decreases} clause in Listing~\ref{algo-compute-root}) and is enough for Dafny to prove termination of \texttt{computeRootUp}.

We establish property 2) by proving a \emph{lemma} (Listing~\ref{proof-computeRoot}): the pre-conditions (\texttt{requires}) of the lemma are the assumptions, and the post-conditions (\texttt{ensures}) the intended property.
The body of the lemma (with a non-interpreted hash function) which provides the machine-checkable proof is available in the \href{https://github.com/ConsenSys/deposit-sc-dafny/blob/50b16f96021368a839f932b5d666729405a305b0/src/dafny/smart/synthattribute/ComputeRootPath.dfy#L85}{computeRootPath.dfy} file. 
This lemma  requires that the tree \texttt{r} is a Merkle tree,
and that the lists \texttt{left} (resp. \texttt{right}) store the values of left (resp. right) siblings of the nodes on a path \texttt{p}.
Moreover, the value at the end of \texttt{p} should be \texttt{seed}. Under these assumptions 
the conclusion (\texttt{ensures}) is that \texttt{computeRootUp} returns the value of 
the root hash of \texttt{r}.


\noindent\begin{minipage}{\linewidth}
  \begin{lstlisting}[language=dafny,caption=Correctness Proof Specification for \texttt{ComputeRootUp}., captionpos=t, label={proof-computeRoot}]
lemma computeRootUpIsCorrectForTree(
  p:seq<bit>,r:Tree<int>,left:seq<int>,right:seq<int>,seed:int) 
  //  size of p is the height of the tree r
  requires |p| == height(r) 
  //  r is a Merkle tree for attribute hash
  requires isCompleteTree(r)          
  requires isDecoratedWith(hash,r)   
  //  the value at the end of the path p in r is seed
  requires seed == nodeAt(p,r).v 
  //  vectors of same sizes
  requires |right| == |left| == |p|   
  // Left and right contain values of left and right siblings of p in r.
  requires forall i :: 0 <= i < |p| ==>
      //  the value of the sibling of the node at p[..i] in r
      siblingValueAt(p,r,i + 1) ==
      //  are stored in left and right  
      if p[i] == 0 then right[i] else left[i]
  //  Main property: computeRootUp computes the hash of the root of r
  ensures r.rootv == computeRootUp(p,left,right,seed)
\end{lstlisting}
\end{minipage}
The proof of lemma \texttt{computeRootUpIsCorrectForTree} requires a few intermediate sub-lem\-mas of moderate difficulty. The main step in the proof is to establish an equivalence between a bottom-up computation \texttt{computeRootUp} and the top-down definition of (attributed) Merkle trees.
All the proofs are by induction on the tree or the path. 
The complete Dafny code for algorithm is available in \href{https://github.com/ConsenSys/deposit-sc-dafny/blob/50b16f96021368a839f932b5d666729405a305b0/src/dafny/smart/synthattribute/ComputeRootPath.dfy}{computeRootPath.dfy} file.

\medskip

Termination for  \texttt{insertValue}  is proved by using a ranking function (decreases clause in Listing~]\ref{algo-compute-sib}).
The functional correctness of \texttt{insertValue} reduces to proving that, assuming \texttt{left} (resp. \texttt{right}) contains the values of the left (resp. right) siblings of the nodes on \texttt{p}, then \texttt{insertValue(p, left, right, seed)} returns the values of the nodes that are left siblings on the successor path.  
The specification of the corresponding lemma is given in Listing~\ref{proof-sib}.
The code for this lemma is in the \href{https://github.com/ConsenSys/deposit-sc-dafny/blob/50b16f96021368a839f932b5d666729405a305b0/src/dafny/smart/paths/NextPathInCompleteTreesLemmas.dfy}{NextPathInCompleteTreesLemmas.dfy} file.
The main proof is based on several sub-lemmas that are not hard conceptually
but cannot be easily discharged using solely the built-in Dafny induction strategies.
They require some intermediate proof hints (verified calculations) to deal with 
all the nodes on the path \texttt{p}.
Note that for this lemma, we require that the leaves are indexed (from left to right)
to be able to uniquely identify each leaf of \texttt{r}.  

\noindent\begin{minipage}{\linewidth}
  \begin{lstlisting}[language=dafny,caption=Correctness Proof Specification for \texttt{ComputeRootUp}., captionpos=t, label={proof-sib}]
lemma insertValuetIsCorrectInATree(
  p: seq<bit>,r:Tree<int>,left:seq<int>,right:seq<int>,seed:T,k :nat)
  //  r is a Merkle tree
  requires isCompleteTree(r)
  requires isDecoratedWith(f, r)
  //  leaves are uniquely indexed from to right
  requires hasLeavesIndexedFrom(r, 0)
  //  k is an index which is not the index of the last leaf
  requires k < |leavesIn(r)| - 1
  requires 1 <= |p| == height(r) 
  // The leaf at index k is the leaf at the end of p
  requires nodeAt(p, r) == leavesIn(r)[k]
  // The value of the leaf at the end of p is seed
  requires seed == nodeAt(p,r).v 
  
  requires |p| == |left| == |right|

  //  Left and right contain the values of the siblings on p
  requires forall i :: 0 <= i < |p| ==>
      siblingAt(take(p,i + 1), r).v == if p[i] == 0 then right[i] else left[i]

  //  A path to a leaf that is not the rightmost one has a zero
  ensures  exists i :: 0 <= i < |p| && p[i] == 0
  //  insertValue computes the values of the left siblings 
  //  of the successor path of `p`.
  ensures forall i :: 0 <= i < |p| && nextPath(p)[i] == 1 ==> 
        computeLeftSiblingOnNextPathFromLeftRight(p,left,right,f,seed)[i] 
        == siblingAt(take(nextPath(p),i + 1),r).v
\end{lstlisting}
\end{minipage}

\paragraph{\bf \itshape Index Based Algorithms.}
The algorithms that implement the DSC do not use a bitvector to encode a path, but rather,
a \emph{counter} that records the number of values inserted so far and the height of the tree.
In order to prove the algorithms actually implemented in the DSC, we first recast the \texttt{computeRootUp} and \texttt{insertValue} algorithms to use a counter and the height $h$ of a tree.
In this step, we use a parameter \texttt{k} that is the index of the next available leaf
where a new value can be inserted. The leaves are indexed left to right from $0$ to $2^{h} - 1$
and hence $k$ is the number of values that have been inserted so far.
It follows that the leaves with indices $k \leq i \leq 2^{h} - 1$ have the default value.
The correspondence between the bitvector encoding of the path to the leaf at index $k$ and the value $k$
is straightforward: the encoding of the path $p$ is the value of $k$ in binary over $h$ bits.
We can rewrite left \texttt{computeRootUp} to use use $k$ and $h$ (\texttt{computeRootUpWithIndex}, Listing~\ref{algo-comp-root-indexed}) 
and prove it computes the same value as \texttt{computeRootUp}.
A similar proof can be established for the \texttt{insertValue} algorithm.
The index based algorithms and the proofs that they are equivalent (compute the same values as) to 
\texttt{computeRootUp} and \texttt{insertValue} are available in the \href{https://github.com/ConsenSys/deposit-sc-dafny/blob/50b16f96021368a839f932b5d666729405a305b0/src/dafny/smart/algorithms/IndexBasedAlgorithm.dfy}{IndexBasedAlgorithm.dfy} file.
Dafny can discharge the equivalence proofs with minimal proof hints using the builtin induction strategies.

\noindent\begin{minipage}{\linewidth}
  \begin{lstlisting}[language=dafny,caption=\texttt{ComputeRootUpWithIndex}., captionpos=t, label={algo-comp-root-indexed}]
computeRootUpWithIndex(
  h:nat,k:nat,left:seq<int>,right:seq<int>,seed:int):int
  requires |left| == |right| == h
  //  the index is in the range of indices for a tree of height h
  requires k < power2(h)
  // Indexed algorithm computes the same value as computeRootUp 
  ensures computeRootUpWithIndex(h,k,left,right,f,seed)  == 
      //  natToBitList(k,h) is the binary encoding of k over h bits
      computeRootUp(natToBitList(k,h),left,right,f,seed)
  //  ranking function
  decreases h 
{
  if h == 0 then seed 
  else if k % 2 == 0 then //  left node 
  computeRootUpWithIndex(h-1,k/2,init(left),init(right),hash(seed,last(right)))
  else // right node
  computeRootUpWithIndex(h-1,k/2,init(left),init(right),hash(last(left),seed))
}
  \end{lstlisting}
  \end{minipage}

  \noindent\begin{minipage}{\linewidth}
    \begin{lstlisting}[language=dafny,caption=Implemented Version of \texttt{computeRootUp}., captionpos=t, label={algo-get-root}]
method get_deposit_root() returns (r:int) 
  //  The result of get_deposit_root_() is the root value of the Merkle tree. 
  ensures r == buildMerkle(values,TREE_HEIGHT,hash).rootv 
{
  //  Store the expected result in a ghost variable.
  //  values is a ghost variable of ther DSC and record all the inserted values
  ghost var e := computeRootUpWithIndex(TREE_HEIGHT,count,branch,zero_hashes,0);
  //  Start with default value for r.
  r := 0;
  var h := 0;
  var size := count;
  while h < TREE_HEIGHT
    //  Main invariant:
    invariant e == 
      computeRootUpWithIndex(
          TREE_HEIGHT - h,size, 
          take(branch,TREE_HEIGHT - h),take(zero_hashes,TREE_HEIGHT - h),r)
  {
    if size % 2 == 1 {
      r := hash(branch[h],r);
    } else {
      r := hash(r,zero_hashes[h]);
    }
    size := size / 2;
    h := h + 1;
  }
}
\end{lstlisting}
  \end{minipage}

\paragraph{\bf \itshape Total Correctness of the Algorithms Implemented in the DSC.}
In this section we present the final proof of (total) correctness for the algorithms
implemented in the DSC (Solidity-like version.)  
Our proof establishes that the imperative versions, with while loops and dynamic memory allocation (for arrays) are correct, terminate and are memory safe.

The DSC is an object and has a state defined by a few variables: \texttt{count} is the number of inserted values (initially zero), \texttt{branch} is a vector that stores that value of the left siblings of the path leading to the leaf at index \texttt{count}, and \texttt{zero\_hashes} is what we previously defined as $z$.
The algorithm that computes the root hash of the Merkle tree in the DSC is \texttt{get\_deposit\_root()}. 
\texttt{get\_deposit\_root()} does not have any \emph{seed} parameter as it computes the
root hash using the default value ($0$). 
The correctness proof of \texttt{get\_deposit\_root()} uses the functional (proved correct) algorithm
\texttt{computeRootUpWithIndex} as an invariant. 
Listing~\ref{algo-get-root} is a simplified version (for clarity) of the
full code available in the  \href{https://github.com/ConsenSys/deposit-sc-dafny/blob/50b16f96021368a839f932b5d666729405a305b0/src/dafny/smart/DepositSmart.dfy}{DepositSmart.dfy} file.

\noindent\begin{minipage}{\linewidth}
  \begin{lstlisting}[language=dafny,caption=The \texttt{deposit} method., captionpos=t, label={algo-deposit}]
method deposit(v:int)   
  //  The tree cannot be full. 
  requires count < power2(TREE_HEIGHT) - 1   
  //  branch and zero_hashes hold the values of the siblings
  requires areSiblingsAtIndex(|values|, 
    buildMerkle(values,TREE_HEIGHT,hash),branch, zero_hashes)  
  //  Correctness property
  ensures areSiblingsAtIndex(|values|, 
    buildMerkle(values,TREE_HEIGHT,hash),branch,zero_hashes)
{
  var value := v;
  var size : nat := count;
  var i : nat := 0;
  //  Store the expected result in e.
  ghost var e := computeLeftSiblingsOnNextpathWithIndex(
    TREE_HEIGHT,old(size),old(branch),zero_hashes,v);
  while size % 2 == 1 
      //  Main invariant:
      invariant e == 
      computeLeftSiblingsOnNextpathWithIndex(
          TREE_HEIGHT - i,size, 
          take(branch,TREE_HEIGHT - i), 
          take(zero_h,TREE_HEIGHT - i),value) + drop(branch,TREE_HEIGHT - i)    
      decreases size 
  {
      value := f(branch[i],value);
      size := size / 2;
      i := i + 1;
  }
  //  0 <= i < |branch| and no there is no index-out-of-bounds error
  branch[i] := value;
  count := count + 1;
  values := values + [v];
}
\end{lstlisting}
\end{minipage}

The algorithm that inserts a value \texttt{v} in the tree is \texttt{deposit(v)} in the
implemented version of the DSC.
Listing~\ref{algo-deposit} is an optimised version of the original algorithm.
The simplification is explained in Section~\ref{sec-findings}. 
The correctness of the algorithm is defined by ensuring that, if at the beginning of the computation
the vectors \texttt{branch} (resp, \texttt{zero\_hashes}) contain values of the left (resp. right) siblings of the path leading to the leaf at index \texttt{count}, then at the end of the computation, after \texttt{v} is inserted,
this property still holds.
The proof of this invariant requires a number of proof hints for Dafny to verify it.
We use the functional version of the algorithm to specify a loop invariant (not provided in Listing~\ref{algo-deposit}).


The termination proof is easy using \texttt{size} as the decreasing ranking function.
However, a difficulty in this proof is memory safety, i.e. to guarantee that the index \texttt{i} used to
access \texttt{branch[i]} is within the range of the indices of \texttt{branch}.  

We have also proved the initialisation functions \texttt{init\_zero\_hashes()} and \texttt{constructor}. 
The full code of the imperative version of the DSC is available in the
\href{https://github.com/ConsenSys/deposit-sc-dafny/blob/50b16f96021368a839f932b5d666729405a305b0/src/dafny/smart/DepositSmart.dfy}{DepositSmart.dfy} file.

\vspace{-1.1em}
\section{Findings and Lessons Learned}\label{sec-findings}

\vspace{-0.79em}


\paragraph{\bf \itshape Methodology.}
In contrast to the previous attempts to analyse the DSC, we have adopted a textbook approach and used 
standard algorithms' design techniques (\eg dynamic programming, refinement, recursion.)
This has several advantages over a direct proof (\eg \cite{formal-inc-merkle-rv}) of the  imperative code including:  
\begin{itemize}
    \item the design of simple algorithms and proofs;
    \item recursive and language-agnostic recursive versions of the algorithms;
    \item new and provably correct simplifications/optimisations.
\end{itemize}

\paragraph{\bf \itshape Algorithmic Considerations.}
Our implementations and formal proofs have resulted in the identification of two previously unknown/unconfirmed optimisations.
First, it is not necessary to initialise the vector of left siblings, \texttt{b}, and the 
  algorithms are correct for any initial value of this vector.

Second, the original  version of the \texttt{deposit} algorithm (which we have proved correct too) 
has the form\footnote{The complete Solidity  source code is freely available on GitHub at \url{https://github.com/ethereum/eth2.0-specs/blob/dev/solidity_deposit_contract/deposit_contract.sol}} given in Listing~\ref{algo-solidity-deposit}.
Our formal machine-checkable proof revealed\footnote{This finding was not uncovered in any of the previous audits/analyses.} that indeed the condition \texttt{C1} is always true and the loop 
always terminates because \texttt{C2} eventually becomes true. 
As witnessed by the comment after the loop in the Solidity code of the DSC, this property was expected but not confirmed,
and the Solidity contract authors did not take the risk to simplify the code.
Our result shows that the algorithm can be simplified to
\texttt{while not(C2) do ... od}. 

\noindent\begin{minipage}{\linewidth}
  \begin{lstlisting}[language=dafny,caption=Solidity Version of the DSC Deposit Function., captionpos=t, label={algo-solidity-deposit}]
deposit( ... )
{
  while C1 do 
    if C2 return; 
  ...
  od
  //  As the loop should always end prematurely with the `return` statement, 
  //  this code should be unreachable. We assert `false` just to be safe.
  assert(false);
} 
\end{lstlisting}
\end{minipage}

This is interesting not only from a safety and algorithmic perspectives, but also because it reduces the computation cost (in gas/Ether) of executing the \texttt{deposit} method.
This simplification proposal is currently being discussed with the DSC developer, however the currently deployed version still uses the non-optimised code.

\paragraph{\bf \itshape Verification Effort.}
The verification effort for this project is 12 person-weeks resulting in $3500$ lines of code and $1000$ lines of documentation.
This assumes familiarity with program verification, Hoare logic and Dafny.
%
%
Table~\ref{tab-stats}, page~\pageref{tab-stats} provides some insights into the code base.

The filenames in \textcolor{green!50!black}{green} are the ones that require the less number of hints for
Dafny to check a proof. 
In this set of files the hints mostly consist of simple \emph{verified calculations} (\eg empty sequence is a neutral element for lists \texttt{[] + l == l + [] == l}.) 
Most of the results on sequences (\texttt{helpers} package) and simplifications of sequences of bits (\texttt{seqofbits} package) are in this category and require very few hints.
This also applies for the proofs\footnote{The file \texttt{CommuteProof.dfy} in this package is not needed for the main proof but was originally used and provides an interesting result, so it is still in code base.} of the \texttt{algorithms} package, \eg proving that the versions using the index of a leaf instead of the binary encoding of a path are equivalent.

\medskip

The filenames in \textcolor{orange}{orange} require some non-trivial proof hints beyond the implicit 
induction strategies built in Dafny. For instance in \textcolor{orange}{NextPathInCompleteTrees.dfy}
and  \textcolor{orange}{PathInCompleteTrees.dfy}, we had to provide several annotations and 
structure for the proofs. This is due to the fact that the proofs involve properties on a Merkle tree
$t_1$ and its \emph{successor} $t_2$ (after a value is inserted) which is a new tree, and on a path $\pi_1$ in $t_1$ and its successor $\pi_2$ in $t_2$. 

\medskip 

The filenames in \textcolor{red}{red} require a lot of hints. 
For the files in the \texttt{synthattribute} package it is mostly calculation steps.
Some steps are not absolutely necessary but adding them reduces the verification time by on order of magnitude (on our system configuration, MacBookPro 16GBRAM).
The hardest proof is probably the correctness of the \texttt{deposit} method in   
\textcolor{red}{DepositSmart.dfy}. The proof requires non trivial lemmas and invariants.
The difficulty stems from a combination of factors: first the while loop of the algorithm (Listing~\ref{algo-deposit}) maintains a constraint between \texttt{size} and \texttt{i}, the latter being used
to access the array elements in \texttt{branch}. 
Proving that 
there is no array-of-bounds error (\ie $i$ is within the size of \texttt{branch})
requires to prove some arithmetic properties. 
Second, the proof of the main invariant (Listing~\ref{algo-deposit}) using the functional specification
\texttt{computeLeftSiblingsOnNextpathWithIndex} is complex and had to be structured around additional lemmas.

\begin{table}[htbp]
    \centering
  \begin{tabular}{lccccc}
    \toprule
                                 src/dafny/\textbf{package}/file.dfy &                          
                                 \#LoC &  Lemmas &  Methods &  \#Doc &  (\#Doc/\#LoC in \%) 
                                 \\
    \midrule
                      \textbf{smart} & & & & & \\   
  
                      \textcolor{red}{DepositSmart.dfy} &              
                      163 &       \textcolor{red}{1} + \textcolor{green!50!black}{1}  &                \textcolor{red}{1} + \textcolor{green!50!black}{3}  &             92 &             56  
                      \\
                      \midrule
                      \textbf{smart/algorithms} & & & & & \\
                      \textcolor{orange}{CommuteProof.dfy} &      
                      73 &         \textcolor{orange}{2} &                0 &             31 &             42 
                      \\
               \textcolor{green!50!black}{IndexBasedAlgorithm.dfy} &      
               96 &         \textcolor{green!50!black}{3} &                \textcolor{green!50!black}{2} &             59 &             61 
               \\
               \textcolor{green!50!black}{MainAlgorithm.dfy} &      
                     66 &         \textcolor{green!50!black}{2} &                0 &             38 &             58 
                     \\
                     \textcolor{green!50!black}{OptimalAlgorithm.dfy} &      
                    24 &         \textcolor{green!50!black}{2} &                0 &             15 &             62 
                    \\
                    \textbf{Sub-total} &  \textbf{259} &       \textbf{\textcolor{orange}{2} + \textcolor{green!50!black}{7} } &   \textbf{\textcolor{green!50!black}{2}} &           \textbf{143} &             \textbf{55} \\ 
                    \midrule
                           \textbf{smart/helpers} & & & & & \\ 
                           \textcolor{green!50!black}{Helpers.dfy} &         
                              51 &         \textcolor{green!50!black}{5} &      \textcolor{green!50!black}{1} &             10 &             20 
                              \\
                              \textcolor{green!50!black}{SeqHelpers.dfy} &        
                         137 &        \textcolor{green!50!black}{10} &                \textcolor{green!50!black}{6} &             34 &             25 
                         \\
                        \midrule
          \textbf{smart/paths} & & & & & \\
          \textcolor{orange}{NextPathInCompleteTrees.dfy} &      
             262 &         \textcolor{orange}{1} + \textcolor{green!50!black}{2} &         \textcolor{green!50!black}{2} &             99 &             38 
             \\
             \textcolor{orange}{PathInCompleteTrees.dfy} &   
                         408 &        \textcolor{orange}{2} + \textcolor{green!50!black}{13}  &                0 &             60 &             15 
                         \\
            \textbf{Sub-total} &  \textbf{670} &       \textbf{\textcolor{orange}{3} + \textcolor{green!50!black}{15}} &   \textbf{\textcolor{green!50!black}{2}} &           \textbf{159} &             \textbf{24} \\ 
            \midrule
                    \textbf{smart/seqofbits} &  & & & & \\  
                    \textcolor{green!50!black}{SeqOfBits.dfy} &    
                             527 &        \textcolor{green!50!black}{19} &                0 &            100 &             19 
                             \\
            \midrule
                   \textbf{smart/synthattribute} &  & & & & \\   
                   \textcolor{orange}{ComputeRootPath.dfy} &  
                   305 &       \textcolor{orange}{2} + \textcolor{green!50!black}{9}  &                0 &            116 &             38 
                   \\
                   \textcolor{green!50!black}{GenericComputation.dfy} &  
                  148 &         \textcolor{green!50!black}{6} &                0 &             75 &             51 
                  \\
                  \textcolor{red}{RightSiblings.dfy} &  
                       210 &         \textcolor{red}{1} + \textcolor{orange}{2} + \textcolor{green!50!black}{2}  &                \textcolor{green!50!black}{1} &             57 &             27 
                       \\
                       \textcolor{red}{Siblings.dfy} &  
                            124 &       \textcolor{red}{1} +  \textcolor{green!50!black}{1}  &                0 &             31 &             25 
                            \\
                            \textcolor{red}{SiblingsPlus.dfy} &  
                        556 &        \textcolor{red}{2} + \textcolor{green!50!black}{2}   &                0 &             52 &              9 
                        \\
                \textbf{Sub-total} &  \textbf{1343} &       \textbf{\textcolor{red}{4} + \textcolor{orange}{4} + \textcolor{green!50!black}{20}} &   \textbf{\textcolor{green!50!black}{1}} &           \textbf{331} &             \textbf{25} \\ 
            \midrule
                    \textbf{smart/trees} & & & & & \\
                    \textcolor{green!50!black}{CompleteTrees.dfy} &     
                               89 &         \textcolor{green!50!black}{8} &            \textcolor{green!50!black}{1}     &             19 &             21   
                               \\
                               \textcolor{green!50!black}{MerkleTrees.dfy} &     
                                208 &         \textcolor{green!50!black}{6} &           \textcolor{green!50!black}{3}    &            101 &             49 
                                \\
                                \textcolor{green!50!black}{Trees.dfy} & 
                                          91 &         \textcolor{green!50!black}{3} &                \textcolor{green!50!black}{5} &             41 &             45 
                                          \\
                    \textbf{Sub-total} &  \textbf{388} &       \textbf{\textcolor{green!50!black}{17}} &   \textbf{\textcolor{green!50!black}{9}} &           \textbf{161} &             \textbf{41} \\ 
                    \midrule
                        \textbf{src/dafny} & &  & & & \\
                            \textbf{TOTAL} &  \textbf{3538} &  ~~~     \textbf{\textcolor{red}{5} 
                            + \textcolor{orange}{9} + \textcolor{green!50!black}{94}} ~~~ &               \textbf{\textcolor{red}{1} + \textcolor{green!50!black}{24}} &           \textbf{1030} &             \textbf{29} 
                            \\
    \bottomrule
    \end{tabular}
    \medskip
    \caption{Dafny Code Statistics. \#Loc (resp. \#Doc) is the number of Lines of Code (resp. Documentation), Lemmas is the number of proofs broken down in difficulty levels, Methods the number of executable methods/function methods. Colour scheme \textcolor{green!50!black}{easy/few proof hints}, \textcolor{orange}{moderate}, \textcolor{red}{hard/detailed proof hints}.}
    \label{tab-stats}
  \end{table}


Overall, almost $90\%$ of the lines of code are (non-executable) proofs, and function definitions used in the proofs.
The verified algorithms implemented in the DSC functional are provided in \texttt{DepositSmart.dfy} and account for less than $10\%$ of the code.


Considering the criticality of the DSC, 12 person-weeks can be considered a moderate effort well worth the investment: the result is an unparalleled level of trustworthiness that can inspire confidence
in the Ethereum platform.  
According to our experts (ConsenSys Diligence) in the verification of Smart Contracts, the size of such an effort is realistic and practical considering the level of guarantees provided. 
The only downside is the level of verification expertise required to 
design the proofs. 

\medskip 

The trust base in our work is composed of the Dafny verification engine (verification conditions generator) and the SMT-solver Z3.



\paragraph{\bf \itshape Dafny Experience.}
Dafny is rather has excellent documentation, support for data structures, functional (side-effect free) and object-oriented programming.  
The automated verification engine has a lot of built-in strategies (\eg induction, calculations) and a good number of specifications are proved fully automatically without providing any hints.
The Dafny proof strategies and constructs that we mostly used are \emph{verified calculations} and \emph{induction}.
The side-effect free proofs seem to be handled much more efficiently (time-wise) than the proofs using mutable data structures.

In the current version we have used the \texttt{autocontracts} attribute for the DSC object which is a convenient way of proving memory safety using a specific 
invariant (given by the \texttt{Valid} predicate). 
This could probably be optimised as Dafny has some support to specify precisely the side-effects using \emph{frames} (based on \emph{dynamic framing}.)
%

Overall, Dafny is a practical option for the verification of mission-critical smart contracts, and a possible avenue for adoption could be to extend the Dafny code generator engine to support Solidity, a popular language for wri\-ting smart contracts for the Ethereum network, or to automatically translate Solidity into Dafny.
We are currently evaluating these options with our colleagues at ConsenSys Diligence, as well as the benefits of our technique to the analysis of other
critical smart contracts.   

\smallskip
The software artefacts including the implementations, proofs, documentation and a Docker container to reproduce the results  are freely available as a GitHub repository at \url{https://github.com/ConsenSys/deposit-sc-dafny}.


\paragraph{\bf \itshape Acknowledgements.} I wish to thank Suhabe Bugrara, ConsenSys Mesh, for helpful dis\-cus\-sions on the Deposit Smart Contract  previous work and the anonymous reviewers of a preliminary version of this paper.   


\bibliography{dafny-sc-bib}

\end{document}